\documentclass[fleqn,pra]{revtex4-2}

\usepackage{amsmath,graphicx}
\usepackage{verbatim}

\begin{document}

\title{Magnetic dipole transitions in the H$_2^+$ ion.}

\author{D.T.~Aznabayev$^{1,2,3}$}
\author{A.K.~Bekbaev$^{2,3}$}
\author{Vladimir I. Korobov$^{1}$}
\affiliation{$^1$Bogoliubov Laboratory of Theoretical Physics, Joint Institute for Nuclear Research, Dubna 141980, Russia,}
\affiliation{$^2$The Institute of Nuclear Physics, Ministry of Energy of the Republic of Kazakhstan, 050032 Almaty, Kazakhstan}
\affiliation{$^3$Al-Farabi Kazakh National University, 050038 Almaty, Kazakhstan}

\begin{abstract}
The magnetic dipole transitions in the homonuclear molecular ion H$_2^+$ are obtained for a range of $v$ and $L$, vibrational and total orbital momentum quantum numbers, respectively. Calculations are performed in the nonrelativistic approximation. Spin consideration is also included.
\end{abstract}

\maketitle

\section{Introduction}

The hydrogen molecular ion is the simplest stable molecule, which may be studied both theoretically and experimentally with high precision.
In recent years laser spectroscopy of the heteronuclear hydrogen molecular ions HD$^+$ achieved spectacular progress \cite{Alighanbari20,Patra20,Kortunov21}. That allowed to get valuable information on fundamental constants such as proton-to-electron mass ratio and set new limits on the possible manifestations of new interactions between hadrons, on the hypothetical "fifth force" \cite{fifth_force21,Schiller23}.

H$_2^+$ is difficult to study experimentally since the electric dipole transitions are strongly suppressed. Nevertheless, the hydrogen molecular ion is of significant interest for metrology, especially as a direct way to measure the proton-to-electron mass ratio, and new experiments are coming to attack the problem using quantum logic spectroscopy \cite{Kienzler23}.

In all the cases it is very important to know the strength of various transitions, which may be induced by the laser light. In our previous studies \cite{quadrupole,E1_forbidden} quadrupole E2 and forbidden E1 transitions for the H$_2^+$ ions were investigated. In the present work we consider M1 transitions for the hydrogen molecular ion H$_2^+$ at low $v$ and $L$.

The M1 transitions in H$_2$ molecule were studied in \cite{PachuckiM1} within the adiabatic approximation, it was found that they are significantly weaker than the quadrupole transitions for low rotational states. The magnetic properties of the H$_2^+$ ion differ from the one of H$_2$ molecule, since the total electron spin $S$ is zero for the case of H$_2$. Still, we come to the similar conclusion for the H$_2^+$ molecular ion. Our calculation is based on \emph{ab initio} three-body variational approximation of the bound state nonrelativistic wave function and a direct evaluation of the orbital magnetic moment operator, see Eq.~(\ref{Eq:gL}) below. No adiabatic approximation is used. Another point, the spin-dependent structure of the transitions, which allow to study the hyperfine spectrum of the H$_2^+$ ion, is carefully considered.

In what follows we assume atomic units: $\hbar=|e|=m_e=1$.

\section{Theory}

Throughout we keep to the following notation: $v$ is the vibrational quantum number, $L$ is the total orbital angular momentum of the nonrelativistic wave function. The spin part is described by the spin operators of two protons, $\mathbf{I}_1$, $\mathbf{I}_2$, and the spin of an electron, $\mathbf{s}_e$, $\mathbf{I}=\mathbf{I}_1+\mathbf{I}_2$ is the total nuclear spin, $\mathbf{F}=\mathbf{I}+\mathbf{s}_e$ is the total spin of the H$_2^+$ ion, and $\mathbf{J}=\mathbf{F}+\mathbf{L}$ is the total angular momentum. Thus the ground "para" state is denoted $(v=0,L=0,I=0,F=1/2,J=1/2)$. Since the M1 transition preserves spatial parity and we will consider the spatial wave function in the nonrelativistic approximation only, we have thus a selection rule for these states: $L\to L'=L$.

The nonrelativistic Hamiltonian in the center of mass frame may be written as:
\begin{equation}\label{H_0}
H_0 
 = \frac{\mathbf{p}_1^2}{2M}+\frac{\mathbf{p}_2^2}{2M}+\frac{\mathbf{p}_e^2}{2m_e}
      -\frac{Z}{r_1}-\frac{Z}{r_2}+\frac{Z^2}{R},
\end{equation}
where $\mathbf{r}_i = \mathbf{r}_e\!-\!\mathbf{R}_i$ and $\mathbf{R} = \mathbf{R}_2\!-\!\mathbf{R}_1$ are electron coordinates relative to nuclei and internuclear position vector in the molecular coordinate notations, $(\mathbf{R}_1,\mathbf{R}_2,\mathbf{r}_e\!\equiv\!\mathbf{R}_3)$ and $(\mathbf{p}_1,\mathbf{p}_2,\mathbf{p}_e\!\equiv\!\mathbf{p}_3)$ are the position vectors and momenta of particles in the center of mass frame, $M=m_p$ is the proton mass, and $Z=1$ is the proton charge.

The transition magnetic moment $\boldsymbol{\mu}_{nn'}$ for a bound system of particles is expressed (see \cite{Landau4}, \S 47):
\begin{equation}\label{Eq:mu_tr}
\boldsymbol{\mu}_{nn'} =
   \left\langle\psi_n\left|\hat{\boldsymbol{\mu}}\right|\psi_{n'}\right\rangle =
   \mu_B\left\langle\psi_n\left|g_L(v,v')\mathbf{L}+g_e\mathbf{s}_e+\frac{g_p}{m_p}\mathbf{I}\right|\psi_{n'}\right\rangle,
\end{equation}
where $\hat{\boldsymbol{\mu}}$ is the operator of the magnetic moment of the bound system, $g_e=-2.002\,319$ is the electron $g$ factor, and $g_p=5.585\,694$ is the proton $g$ factor \cite{CODATA18}, $\mu_B=|e|\hbar/(2m_ec)$ is the Bohr magneton, $g_L(v,v')$ is the orbital $g$-factor of the transition:
\begin{equation}\label{Eq:gL}
g_L(v,v') =
   \frac{1}{\mu_B\left\langle L \left\| \mathbf{L} \right\| L \right\rangle}\:
   \sum_{a=1}^3\left\langle\psi_n\left\|\frac{z_a}{m_a}\left[\mathbf{R}_a\times\mathbf{p}_a\right]\right\|\psi_{n'}\right\rangle,
\end{equation}
here the sum is over all the particles of the system, and $z_a$ and $m_a$ are corresponding charges and masses of the particles.

When $n\ne n'$ the initial and final state functions are orthogonal and the spin terms in (\ref{Eq:mu_tr}) are exactly zero. Thus the transition amplitudes are determined solely by $g_L$, the orbital g-factor of the transition. The spin dependent transition amplitudes between hyperfine states are then expressed:
\begin{equation}\label{Eq:T_HFS}
\left\langle vIFLJ\|\hat{\boldsymbol{\mu}}\|v'IF'LJ'\right\rangle =
   \delta_{FF'}(-1)^{J+F+L+1}\sqrt{(2J\!+\!1)(2J'\!+\!1)}
   \left\{\begin{matrix}
      L & 1 & L \\ J' & F & J
   \end{matrix}\right\}
   \left\langle vL\|\hat{\boldsymbol{\mu}}\|v'L\right\rangle\,,
\end{equation}
and we have an additional selection rule for the hypefine transition lines: $F\to F'=F$.

The transition probability for spontaneous emission of a magnetic dipole photon from state $n$ to state $n'$ is expressed as follows
\begin{equation}\label{Eq:Einstein}
A_{nn'}
 = \frac{1}{\tau}\,\frac{4\alpha^3}{3}\,w_{nn'}^3\,
   \frac{\left|\left\langle\psi_n
      \left\|\hat{\boldsymbol{\mu}}\right\|
   \psi_{n'}\right\rangle\right|^2}{2J_n+1}
 = \frac{1}{\tau}\,\frac{\alpha^5}{3}\,w_{nn'}^3\,
    \frac{g_L^2(v,v')}{2J_n+1},
\end{equation}
here $w_{nn'}=E_n-E_{n'}$ is the transition frequency, and $\tau$ is a unit of time in atomic units: $\tau=2.41888\times10^{-17}$ s.

If $v=v'$, transitions occur between hyperfine states of the same vibrational level and all the terms in (\ref{Eq:mu_tr}) contribute to the transition strength. For the "pure" HFS states (for definition see \cite{quadrupole}, Eq.~(10)) the matrix of the magnetic moment is expressed by
\begin{equation}\label{Eq:HFS}
\boldsymbol{\mu}_{nn'} =
   \mu_B\left\langle vIFLJ\left\|g_L\mathbf{L}\!+\!g_e\mathbf{s}_e\!+\!\frac{g_p}{m_p}\mathbf{I}\right\|vI'F'LJ'\right\rangle,
\end{equation}
then for the real hyperfine eigenstates the matrix is modified by orthogonal transformation, which connects "pure" and real HF eigenstates. The matrix elements for particular terms in (\ref{Eq:HFS}) may be evaluated using 6-$j$ symbols as is shown in the Appendix, Eq.~(\ref{Eq:reduced}).

\begin{table}
\begin{center}
\begin{tabular}{c@{\hspace{3mm}}|@{\hspace{3mm}}c@{\hspace{6mm}}c@{\hspace{6mm}}c@{\hspace{3mm}}|@{\hspace{3mm}}c@{\hspace{6mm}}c@{\hspace{6mm}}c}
\hline\hline
 & \multicolumn{3}{c}{$L=1$} & \multicolumn{3}{c}{$L=2$} \\
\hline
$v\!\to\!v'$ & $v=0$ & $v=1$ & $v=2$ & $v=0$ & $v=1$ & $v=2$ \\
\hline
$v'=0$ & 5.0113($-$4) &   --    &     --       & 5.0095($-$4) &      --      &    --   \\
1 & 7.6094($-$6) & 4.9606($-$4) &     --       & 7.6324($-$6) & 4.9587($-$4) &    --   \\
2 & 6.5530($-$7) & 1.1257($-$5) & 4.9060($-$4) & 6.5775($-$7) & 1.1291($-$5) & 4.9039($-$4) \\
3 & 1.2180($-$7) & 1.1844($-$6) & 1.4421($-$5) & 1.2221($-$7) & 1.1892($-$6) & 1.4465($-$5) \\
4 & 3.4906($-$8) & 2.4674($-$7) & 1.7594($-$6) & 3.5029($-$8) & 2.4765($-$7) & 1.7670($-$6) \\
5 & 1.2655($-$8) & 7.7464($-$8) & 3.9817($-$7) & 1.2705($-$8) & 7.7754($-$8) & 3.9977($-$7) \\
6 & 5.3465($-$9) & 3.0371($-$8) & 1.3402($-$7) & 5.3693($-$9) & 3.0498($-$8) & 1.3456($-$7) \\
7 & 2.5135($-$9) & 1.3718($-$8) & 5.6022($-$8) & 2.5417($-$9) & 1.3783($-$8) & 5.6268($-$8) \\
8 & 1.3084($-$9) & 6.8817($-$9) & 2.6810($-$8) & 1.3164($-$9) & 6.9173($-$9) & 2.6953($-$8) \\
9 & 7.3091($-$10)& 3.7398($-$9) & 1.4170($-$8) & 7.3522($-$10)& 3.7732($-$9) & 1.4225($-$8) \\
10& 4.3382($-$10)& 2.2204($-$9) & 7.9503($-$9) & 4.3783($-$10)& 2.2064($-$9) & 8.0930($-$9) \\
\hline\hline
\end{tabular}
\end{center}
\caption{Orbital $g$-factor for transition, $g_{L}(v,v')$ for the orbital states $L=1$ and $L=2$. ($a(b)=a\!\times\!10^b$).}\label{T:L1}
\end{table}

\begin{table}
\begin{center}
\begin{tabular}{c@{\hspace{6mm}}c@{\hspace{6mm}}c@{\hspace{6mm}}c@{\hspace{6mm}}c}
\hline\hline
$v\to v'$ & $L=1$ & $L=2$ & $L=3$ & $L=4$ \\
\hline
$0\!\to\!1$ & 7.6094($-$6)  & 7.6324($-$6)  & 7.6670($-$6)  & 7.7132($-$6)  \\
$0\!\to\!9$ & 7.3091($-$10) & 7.3522($-$10) & 7.4243($-$10) & 7.5216($-$10) \\
\hline\hline
\end{tabular}
\end{center}
\caption{The dependence of $g_L$ on the orbital momentum $L$ for two vibrational transitions.}\label{T:2}
\end{table}

\begin{figure}
\caption{Dependence of $g_L$ for various $v\to v'$ transitions. $L=1$.}
\includegraphics[width=0.5\textwidth]{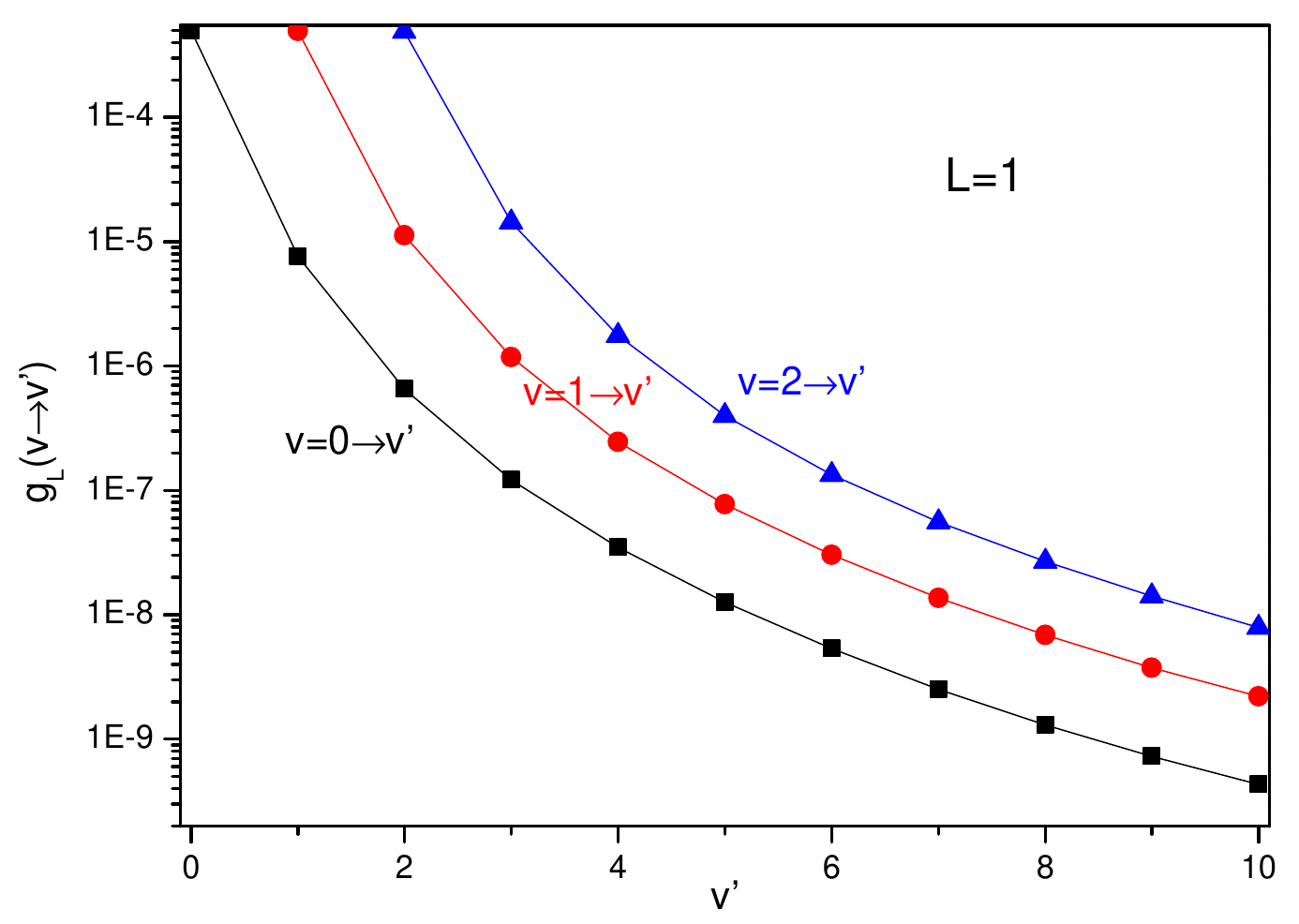}
\end{figure}

\section{Results and discussion}

Numerical calculations were based on the "exponential" variational expansion \cite{Korobov00}. By averaging Eq.~(\ref{Eq:gL}) over the spatial degrees of freedom, we obtained the orbital $g$-factors and, eventually, the Einstein coefficients $A_{nn'}$ for the spontaneous emission of a photon from the state $n=(v,L,F,J)$ to the state $n'=(v',L',F',J')$, see Eq.~(\ref{Eq:Einstein}).

In Table \ref{T:L1} the results of our numerical calculations for $L\!=\!1$ and $L\!=\!2$ transitions are presented. The transition probability calculated from the orbital $g$-factors for the $(L=1,v=1)\to(1,0)$ transition is $A_{nn'}^{M1}=5.45\times10^{-12}$ s$^{-1}$, which must be compared with the quadruple transition probability for the same transition $A_{nn'}^{E2}=2.65\times10^{-7}$ s$^{-1}$ \cite{quadrupole}. Thus, the M1 transition is five orders of magnitude weaker than the quadrupole one.

Variation of the orbital $g$-factor  with a change of $L$ is expectedly small due to the adiabatic nature of the system, and this is confirmed numerically in Table \ref{T:2}. Since the $L$ dependence is small we do not present calculations for higher $L$ states. On the other hand, the orbital magnetic moment is proportional to $\sqrt{L(L+1)(2L+1)}$ and transition probability $A_{nn'}$ increases with $L$, similarly to the case of the molecule H$_2$ (see Fig.~1, Ref.~\cite{PachuckiM1}) and at some $L>20$ the M1 transition becomes dominant.

On Figure 1 three series of vibrational transitions for the $L=1$ states are shown.

In conclusion, we obtained the M1 transition strenths for the hydrogen molecular ion H$_2^+$. To our knowledge this is the first systematic study of the M1 transitions for this molecule. Along with our previous calculations of the quadruple E2 transitions and forbidden dipole E1 transitions it completes the study of the transition rates, which are necessary for planning future experiments and allow to estimate the laser intensity required for precision spectroscopy of the H$_2^+$ ion.

\section*{Acknowledgements}

D.Aznabayev acknowledges the support of Ministry of Science and Higher Education of the Republic of Kazakhstan under Grant No. AP19175613,
V.K. acknowledges the support of the Russian Science Foundation under Grant No. 23-22-00143.

\appendix

\section*{Reduced matrix elements for the spin and orbital angular momentum operators.}

In this section we will apply the formula for the tensor product of irreducible tensor operators \cite{messiah}.

Let $\mathbf{V}_{k\mu}\equiv\{\mathbf{T}_{k_1}\otimes\mathbf{U}_{k_2}\}_{k\mu}$, where $\mathbf{T}_{k_1}$ acts on subsystem 1 with angular moment $\mathbf{J}_1$ and $\mathbf{U}_{k_2}$ acts on subsystem 2 with angular momentum $\mathbf{J}_2$. Then
\begin{equation}\label{Eq:tensor}
\left\langle v_1v_2J_1J_2J\|\mathbf{V}_k\|v_1'v_2'J_1'J_2'J'\right\rangle
   = \Pi_{JJ'k}
   \left\{\begin{matrix}
      J'_1 & J'_2 & J' \\ k_1 & k_2 & k \\ J_1 & J_2 & J
   \end{matrix}\right\}
   \left\langle v_1J_1\|\mathbf{T}_{k_1}\|v_1'J_1'\right\rangle
   \left\langle v_2J_2\|\mathbf{U}_{k_2}\|v_2'J_2'\right\rangle.
\end{equation}

The reduced matrix elements for operators $\mathbf{I}$, $\mathbf{s}_e$ and $\mathbf{L}$ in the basis of the "pure" hyperfine states are:
\begin{equation}\label{Eq:reduced}
\begin{array}{@{}l}
\displaystyle
\left\langle IFLJ\|\mathbf{I}\|IF'LJ'\right\rangle
   = \Pi_{FF'JJ'}
   (-1)^{I+s_e+L+J'+1}
   \left\{\begin{matrix}
      I & \!1\! & I \\ F' & \!s_e\! & F
   \end{matrix}\right\}
   \left\{\begin{matrix}
      F & \!1\! & F' \\ J' & \!1\! & J
   \end{matrix}\right\}
   \left\langle I\|\mathbf{I}\|I\right\rangle,
\\[4mm]\displaystyle
\left\langle IFLJ\|\mathbf{s}_e\|IF'LJ'\right\rangle
   = \Pi_{FF'JJ'}
   (-1)^{I+s_e+L+J'+1}
   \left\{\begin{matrix}
      s_e & \!1\! & s_e \\ F' & \!I\! & F
   \end{matrix}\right\}
   \left\{\begin{matrix}
      F & \!1\! & F' \\ J' & \!L\! & J
   \end{matrix}\right\}
   \left\langle s_e\|\mathbf{s}_e\|s_e\right\rangle,
\\[4mm]\displaystyle
\left\langle IFLJ\|\mathbf{L}\|IF'LJ'\right\rangle
   = \delta_{FF'}\Pi_{JJ'}
   (-1)^{F+L+J+1}
   \left\{\begin{matrix}
      L & \!1\! & L \\ J' & \!F\! & J
   \end{matrix}\right\}
   \left\langle L\|\mathbf{L}\|L\right\rangle.
\end{array}
\end{equation}
where $\Pi_{S}=\sqrt{2S\!+\!1}$ and $\langle S\|\mathbf{S}\|S \rangle=\sqrt{S(S\!+\!1)(2S\!+\!1)}$ for any angular momentum operator $\mathbf{S}$.

\clearpage

\end{document}